\documentstyle[preprint,aps,epsf,floats]{revtex}
\begin{document}
\tighten

\def\bfl{{\bbox \ell}}
\def\bull{\vrule height .9ex width .8ex depth -.1ex}
\def\Dslash{ {D\hskip-0.6em /} }
\def\MeV{{\rm MeV}}
\def\GeV{{\rm GeV}}
\def\Tr{{\rm Tr\,}}
\def\D{{\Delta}}
\def\Ds{{\Delta_6}}
\def\a{{\alpha}}
\def\b{{\beta}}
\def\c{{\gamma}}
\def\d{{\delta}}
\def\m{{\mu}}
\def\M{{\cal M}}
\def\C{{\cal C}}
\def\slash{{\!\not\!}}
\def\nrcpt{NR\raise.4ex\hbox{$\chi$}PT\ }
\def\ket#1{\vert#1\rangle}
\def\bra#1{\langle#1\vert}
\def\ltap{\ \raise.3ex\hbox{$<$\kern-.75em\lower1ex\hbox{$\sim$}}\ }
\def\gtap{\ \raise.3ex\hbox{$>$\kern-.75em\lower1ex\hbox{$\sim$}}\ }
\newcommand{\gsim}{\raisebox{-0.7ex}{$\stackrel{\textstyle >}{\sim}$ }}
\newcommand{\lsim}{\raisebox{-0.7ex}{$\stackrel{\textstyle <}{\sim}$ }}

\def\Journal#1#2#3#4{{#1} {\bf #2}, #3 (#4)}

\def\NCA{\em Nuovo Cimento}
\def\NIM{\em Nucl. Instrum. Methods}
\def\NIMA{{\em Nucl. Instrum. Methods} A}
\def\NPB{{\em Nucl. Phys.} B}
\def\NPA{{\em Nucl. Phys.} A}
\def\PLB{{\em Phys. Lett.}  B}
\def\PRL{\em Phys. Rev. Lett.}
\def\PRD{{\em Phys. Rev.} D}
\def\PRC{{\em Phys. Rev.} C}
\def\PRA{{\em Phys. Rev.} A}
\def\PR{{\em Phys. Rev.} }
\def\ZPC{{\em Z. Phys.} C}
\def\PREP{{\em Phys. Rep.}  }
\def\ANN{{\em Ann. Phys.} }
\def\SCI{{\em Science} }
\def\CJP{{\em Can. J. Phys.}}

\preprint{\vbox{
\hbox{DOE/ER/40561-91-INT00}
\hbox{NT@UW-00-06}
}}
\bigskip
\bigskip

\title{Renormalization Group Improved Gap Equation for
Color~Superconductors}
\author{\bf Silas R. Beane$^a$,  
Paulo F. Bedaque$^b$, and
Martin J. Savage$^{a,c}$ }

\vspace{1cm}

\address{$^a$ Department of Physics 
\\ University of Washington, Seattle, WA 98195-1560}

\vspace{1cm}

\address{$^b$ Institute for Nuclear Theory
\\ University of Washington, Seattle, WA 98195-1560}

\vspace{1cm}

\address{$^c$ Jefferson Laboratory\\ 12000 Jefferson Avenue, Newport News, 
VA 23606}
\vspace{1cm}
\address{\tt sbeane, bedaque, savage@phys.washington.edu}

\maketitle

\begin{abstract}
The renormalization group is used to resum leading logarithmic
contributions
of the form $\alpha_s^{n+1}\beta_0^n\log^n\left(\D/\mu\right)$
to the gap equation appropriate
for high density QCD.
The scale dependence of the strong coupling constant $\alpha_s$ increases the 
gap by  a factor of 
$\exp\left({33\over 16}\left({\pi^2\over 4}-1\right)\right)
\sim 20$ in the small coupling limit.

\end{abstract}

\vfill\eject

The present efforts towards understanding QCD at extremely high
density~\cite{alford1}-\cite{ALL00} are an important step toward
understanding QCD at moderate densities which may one day be
accessible to experimental study.  For particular combinations of
quark colors and flavors at high densities, it is likely that a
superconducting gap breaks color and flavor symmetries in interesting
ways.  Although this symmetry breaking is nonperturbative, it occurs
when QCD is weakly coupled, and therefore perturbative QCD (pQCD) can
be used to derive properties of the superconducting phase.  The most
basic property is the size of the gap itself.  In beautiful work,
Son\cite{SON1} realized that the gap in pQCD is dominated by
magnetic gluon exchange, regulated in the infrared by dynamic
screening.  He arrived at an estimate of the gap by renormalization
group (RG) arguments, and found $\D\sim c g_s^{-5}
\exp\left(-{3\pi^2\over\sqrt{2} g_s}\right)$ in the small coupling
limit, where $g_s$ is the strong coupling constant evaluated at a
scale of order the chemical potential, $\mu$, and $c$ is a numerical
coefficient.  This surprising result was also obtained directly from
the Schwinger-Dyson (SD) equations\cite{SchaferWilczek}.  Further,
numerical solution of the gap equation\cite{SchaferWilczek} with
particular boundary conditions yields values of the gap that are within
an order of magnitude of the small coupling result for the same
boundary conditions.

In order to refine the perturbative estimate of the gap, the
coefficient $c$ must be determined.  At this one-loop order in the SD
equations there are several important contributions to $c$ 
[e.g. \cite{rock} ], one being
the ambiguity in the scale at which the strong coupling constant $g_s$
is evaluated.  Working in an effective theory appropriate for momentum
transfers below $2\mu$, it has been suggested that choosing a
renormalization scale $\lambda=\mu$ (mid way between $2\mu$ and $0$)
provides an estimate of the contribution to the running of
$g_s$\cite{SchaferWilczek}.  While plausible, in order to determine
$c$ a somewhat more rigorous estimate is required, and in fact we will
show that changing the scale of $g_s$ in the solution of the gap
equation
 does not capture the largest
contribution.  The scale ambiguity in the gap equation, as discussed
in
\cite{SON1,SchaferWilczek}, will be reduced if computations are performed
at higher orders in the $g_s$ expansion,  
i.e. ${\cal O}(\alpha^2_s)$.
For many processes, 
an estimate of the scale can be made by computing
terms of the form $\alpha^2\beta_0$, 
($\beta_0=11-{2\over 3} n_f$ in vacuum, $n_f$ being the number of
dynamical quark flavors)
from  the gluon vacuum polarization diagrams arising at two-loops
\cite{BLM}.  
Introduced by Brodsky, Lepage and Mackenzie (BLM), this scale setting procedure
has been used extensively for several different processes, including the 
inclusive decay of heavy quarks, e.g. \cite{LSW}.
However, 
there is a problem applying this technique to the gap equation
because $\log\D$ is an expansion in $g_s$ and not $\alpha_s$.
In this work we will compute and resum contributions of the form
$\alpha^{n+1}\beta_0^n\log^n\left(\D/\mu\right)$ 
to the gap equation for $\D$
using the renormalization group.

\begin{figure}[t]
\centerline{{\epsfxsize=5.0in \epsfbox{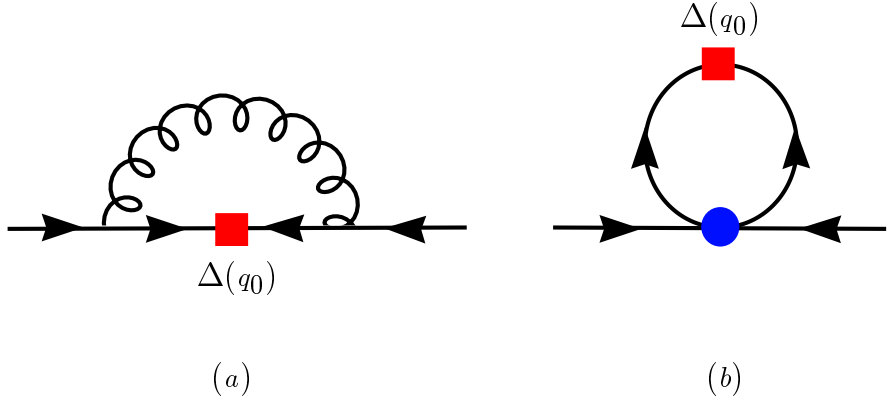}} }
\noindent
\caption{\it
The leading diagrams contributing to the gap equation.
The solid square denotes the gap, $\Delta$, while the 
solid circle denotes an insertion of the counterterm.
}
\label{fig:gap}
\vskip .2in
\end{figure}

Explicit construction of the 
effective field theory with which we work,
appropriate for momentum scales below 
$2\mu$, can be found in papers by  Hong\cite{hong}.
Neglecting higher order  
coupling between the SD equations relating the gluon and 
quark two-point functions, only the graphs 
shown in Fig.~(\ref{fig:gap}) need be calculated
in order to resum the leading logarithmic contributions.
%
Regularizing the perpendicular gluon momentum $k_\perp$ with
dimensional regularization and minimal subtraction one finds that
at $\lambda\sim 2\mu$ the gap equation is
\begin{eqnarray}
\D (p_0) & = & 
-{1\over 24\pi^2}\int_{-\infty}^\infty\ dq_0
{\D (q_0)\over\sqrt{ q_0^2 + \D^2(q_0)}}\ 
\left[ g_s^2(\lambda) \left( {2\over 3} \log\left({\lambda_M^2
        |p_0-q_0|\over\lambda^3}\right)
\ +\  \log\left({\lambda_E^2\over\lambda^2}\right)
\right)
\ +\ \C (\lambda)\right].
\label{eq:gaphigh}
\end{eqnarray}
We have neglected the $\left(p_0-q_0\right)^2$ 
terms in the  gluon propagators as they constitute higher order
corrections in the effective field theory.
In order to obtain eq.~(\ref{eq:gaphigh}), we have used the small $q_0$ limit
of the Landau damping term,
\begin{eqnarray}
\lambda_M^2 & = & {n_f g_s^2 \mu^2\over 8\pi}
\ \ ,\ \ 
\lambda_E^2\ =\ {n_f g_s^2 \mu^2\over 2\pi^2}
\ \ .
\label{eq:screen}
\end{eqnarray}
The ``constant'' $\C$ is the coefficient of a four-quark operator in the
effective theory whose value is determined by matching QCD to the effective
field theory.
In order to absorb the scale dependence of the one-loop graphs, $\C$ has
precisely
equal and opposite scale dependence.
At higher orders in the $1/\mu$ and $g_s$ expansion
four-quark operators will enter with scale and gauge dependence that precisely
compensates the scale and gauge dependence of the loop graphs in the effective
theory.
With no further mention of the gauge dependence of $\C$, we exploit the 
scale dependence of $\C$ to resum all leading log contributions to the 
gap equation.
In what follows we neglect the scale dependence of the $g_s$'s that appear 
in eq.~(\ref{eq:screen}), and assume that they are evaluated at $2\mu$,
as variations from these terms are higher order effects.
In fact, it is simpler to separately evolve the contribution from electric and 
magnetic gluons by setting $\C=\C^{(E)}+\C^{(M)}$.
The assertion that $\Delta$ is scale independent leads to a differential
equation for the $\C^{(i)}$, 
\begin{eqnarray}
\lambda {d\C^{(i)}\over d\lambda} & = & \gamma^{(i)}\  g_s^2 
\ \ \ ,
\label{eq:RG}
\end{eqnarray}
where $ \gamma^{(i)}=2$ for both $i=E,M$. 
Straightforward solution of this differential equation leads to a 
RG improved gap equation, where we chose $\lambda=\lambda_E$ for the
electric and $\lambda=\lambda_M$ for the magnetic gluons)
\begin{eqnarray}
& & \D (p_0)\ =\ 
-\int_{-\infty}^\infty\ dq_0
{\D (q_0)\over\sqrt{ q_0^2 + \D^2(q_0)}}\
\nonumber\\
& &  
\left[ {\C (2\mu)\over 24\pi^2}
\ +\ 
\log\left(
\left({\alpha_s (2\mu)\over\alpha_s (\lambda_E)}
\right)^{\gamma^{(E)}\over 3\beta_0}
\left({\alpha_s (2\mu)\over\alpha_s (\lambda_M^{2/3} |p_0-q_0|^{1/3})}
\right)^{\gamma^{(M)}\over 3\beta_0}
\right)
\right]
\ \ \ .
\label{eq:gaplow}
\end{eqnarray}
An important point to note here is that the gap equation does depend upon 
the chemical potential $\mu$, however, it is independent of the renormalization
scale $\lambda$ in the effective field theory.
While $\Delta (p_0)$ is symmetric under $p_0\rightarrow -p_0$,
the integrand is not symmetric under $q_0\rightarrow -q_0$.
In the region where both electric and magnetic gluons are effectively massless,
$\lambda_E < \lambda < 2 \mu$,
the evolution of the strong coupling is determined by $\beta_0=11$, 
which can be seen by considering the gluon three-point
function and making use of gauge invariance.
However, the $\beta$-function in the region
below $\lambda_E$, where the electric gluons are not dynamical, is yet to be 
computed; in eq.~(\ref{eq:gaplow}) and throughout 
this work we will assume that the magnetic and eletric $\beta_0$'s are 
the same.


Proceeding along lines similar to Son\cite{SON1},  
a second order differential equation for $\D (p_0)$ can be obtained, 
and solved for given boundary conditions.
However, in contrast to Son who sets up a differential equation in the variable
$\log\left(p_0\right)$, we use the variable 
$\alpha_s \left(\lambda_M^{2/3} p_0^{1/3}\right)$, which
at leading order recovers the Son result, and enables us to
include the higher order contributions from the evolution of $g_s$.

Assuming that 
$q_0\gg\D$, we can write the gap equation in eq.~(\ref{eq:gaplow}) as 
\begin{eqnarray}
\D (\alpha_p) & = &  -{6\pi\over\beta_0}
\int_0^\infty d\alpha_q
{\D (\alpha_q)\over\alpha_q^2}
\left[ {\C (2\mu)\over 12\pi^2}
\ +\ 
\log\left(
\left[{\alpha_s (2\mu)\over\alpha_s (\lambda_E)}
\right]^{2\gamma^{(E)}\over 3\beta_0}
\left[{\alpha_s (2\mu)\over\alpha_{p-q}}
\right]^{\gamma^{(M)}\over 3\beta_0}
\left[{\alpha_s (2\mu)\over\alpha_{p+q}}
\right]^{\gamma^{(M)}\over 3\beta_0}
\right)
\right],
\label{eq:gapalpha}
\end{eqnarray}
where $\alpha_k = \alpha_s (\lambda_M^{2/3} |k_0|^{1/3})$.
Using the somewhat brutal approximation of \cite{SON1},

\begin{eqnarray}
\alpha_s (|p_0-q_0|) & = & 
\theta (\alpha_s (q_0)-\alpha_s (p_0)) \alpha_s (p_0)
\ +\ 
\theta (\alpha_s (p_0)-\alpha_s (q_0)) \alpha_s (q_0)
\ \ \ ,
\label{eq:brutality}
\end{eqnarray}
we are able to convert the integral equation into the differential equation

\begin{eqnarray}
\alpha_p^3 \ { d^2\D\over d\alpha_p^2}
\ +\ 
\alpha_p^2 \ { d\D\over d\alpha_p}
\ +\ 
{4\pi\gamma^{(M)}\over\beta_0^2}\ \D & = & 0
\ \ \ .
\label{eq:diff}
\end{eqnarray}
The general solution to this differential equation is 
\begin{eqnarray}
\D (\alpha_p) & = & 
A_1\ J_0\left({4\over\beta_0}\sqrt{\pi\gamma^{(M)}\over\alpha_p}\right)
\ +\ 
A_2\ Y_0\left({4\over\beta_0}\sqrt{\pi\gamma^{(M)}\over\alpha_p}\right)
\ \ \ ,
\label{eq:desoln}
\end{eqnarray}
where $J_0 (x)$ and $Y_0(x)$ are Bessel and Neumann functions,
respectively.
In the limit where $\alpha_\D$ 
is small, a self consistent solution where 
$g_s \log\left(\D/\mu\right)\sim~1$ can be found.
Taking the asymptotic limit of the Bessel functions in eq.~(\ref{eq:desoln}),
\begin{eqnarray}
\D (\alpha_p) & = & A\ \alpha_p^{1/4}
\ \sin\left[ { 4\over \beta_0}\sqrt{\pi\gamma^{(M)}\over \alpha_p} 
+ \phi\right]
\ \ \ ,
\label{eq:asym}
\end{eqnarray}
where $A$ and $\phi$ are constants.
As expected this reduces to the solution found by Son\cite{SON1}
when the argument of the Sine function is expanded in powers of $\log p_0$,
but with the advantage that we have been able to resum the leading logs
directly.

In order to determine $A$ and $\phi$ we return to the integral equation
in eq.~(\ref{eq:gapalpha}).
Given the form of the gap in eq.~(\ref{eq:asym}), it is clear that the 
integral on the right hand side of eq.~(\ref{eq:gapalpha}) is 
divergent, as $\alpha\rightarrow 0$, for any value of $\phi$, and 
hence needs to be regulated in the ultra-violet.
While unrelated, similar integral equations appear in 
effective field theory studies of the 
three-body problem\cite{threebod}.
Efforts to dimensionally regulate such integrals have failed up 
to this point, however, cut-off regulation has been used 
successfully\cite{threebod}.
Therefore, we will introduce a cut-off, $\alpha_\Lambda$, on the
longitudinal direction to define 
the integral equation and consequently, the countertern $C(2\mu)$
becomes dependent on this cutoff in the longitudinal direction
 $C(2\mu)=C(2\mu,\alpha_\Lambda)$
Using the approximation in eq.~(\ref{eq:brutality}) and 
$\gamma^{(E)}=\gamma^{(M)}=2$,
eq.~(\ref{eq:gapalpha}) becomes (including the infrared cut-off defined by 
$\Delta$, $\alpha_\D$)
\begin{eqnarray}
\D (\alpha_p) & = & 
{8\pi\over\beta_0^2}
\left[\ 
\int_{\alpha_\Lambda}^{\alpha_p}
\  d\alpha_q
{\D (\alpha_q)\over\alpha_q^2}
\log\left({\alpha_q\over\overline{\alpha}}\right)
\ +\ 
\log\left({\alpha_p\over\overline{\alpha}}\right)
\int_{\alpha_p}^{\alpha_\D}
\  d\alpha_q
{\D (\alpha_q)\over\alpha_q^2}
\right.\nonumber\\ 
&\ \ \ &
\left.
-\frac{\beta_0}{16\pi^2}(C(2\mu,\alpha_\Lambda))-C(2\mu,\bar\alpha)) 
\int_{\alpha_\Lambda}^{\D}
\  d\alpha_q
{\D (\alpha_q)\over\alpha_q^2}
\ \right],
\label{eq:gapcutoff}
\end{eqnarray}
where
\begin{eqnarray}
\overline{\alpha} & = & 
{\alpha_s^2 (2\mu)\over\alpha_s (\lambda_E)}
\exp\left({\beta_0 C(2\mu,\bar\alpha) \over 16\pi^2}\right)
\ \ \ .
\label{eq:abar}
\end{eqnarray}

It is clear from eq.~(\ref{eq:gapcutoff}), that $\D (\overline{\alpha})=0$,
when $\alpha_\Lambda=\overline{\alpha}$.
As the result is required to be independent of  $\alpha_\Lambda$, we find
that,
in the small $\alpha_\Lambda$ limit
\begin{eqnarray}
C(\alpha_\Lambda)=C(\bar\alpha)+
16\pi^3 \left(\sin(\frac{4}{\beta_0}\sqrt{\frac{2\pi}{\bar\alpha}})
-\sin(\frac{4}{\beta_0}\sqrt{\frac{2\pi}{\alpha_\Lambda}})\right)
\frac{1}{\cos(\frac{4\sqrt{2 \pi}}{\beta_0}
(\frac{1}{\sqrt{\alpha_\Lambda}-\sqrt{\bar\alpha}}))}
-\frac{16\pi^2}{\beta_0}\log(\frac{\bar\alpha}{\alpha_\Lambda})
\label{eq:ct}
\end{eqnarray}
It is important to notice the dependence of $C(2\mu,\alpha_\Lambda)$ on the
$\alpha_\Lambda$.  
Unlike the counterterms in perturbative effective field theories,
this counterterm changes rapidly with scale (except near
$\alpha_\Lambda=\overline{\alpha}$), 
and is periodic, as  found in
three-body physics \cite{threebod}.  In the small coupling limit,
$C(2\mu,\alpha_\Lambda )\sim C(2\mu,\overline{\alpha} )\sim g_s^2 (2\mu)$.

As the solution to the gap equation 
is independent of $\alpha_\Lambda$, by construction, 
we choose to work
at $\alpha_\Lambda=\overline{\alpha}$.
We find
\begin{eqnarray}
\D (\alpha_p) & = & A\ \alpha_p^{1/4}
\ \sin\left[ { 4\sqrt{2\pi}\over \beta_0}
\left(
{\sqrt{\alpha_s(\lambda_E)}\over \alpha_s(2\mu)}
\exp\left({\tilde C \beta_0\over 32\pi^2}\right)-
{1\over\sqrt{\alpha_p}}
\right) 
\right]
\ \ \ .
\label{eq:asymc}
\end{eqnarray}
{}From eq.~(\ref{eq:gapcutoff}) it is clear that
$\alpha_p=\alpha_\D$ when ${d\Delta\over d\alpha_p}=0$,
which leads to 
\begin{eqnarray}
\D (\D) & = & {1024\sqrt{2} \pi^4\over n_f^{5/ 2} g_s^5}\ 
\exp\left(-{3\pi^2\over \sqrt{2}  g_s}\right)\ 
\exp\left({3\beta_0\over 16}\left[{\pi^2\over 4}-1\right]
\right)\ 
\exp\left(-d\right)\ 
\ \ \ ,
\label{eq:gapcrap}
\end{eqnarray}
where $g_s= g_s(2\mu)$, and
\begin{eqnarray}
d & = & {3  C(2\mu,\bar\alpha) \over 2 g_s^2}
\ \ \ ,
\label{eq:ddef}
\end{eqnarray}
is expected to be a number of order unity.
The first two factors constitute the 
result obtained by \cite{SON1} and 
\cite{SchaferWilczek}.
The third factor
$\exp\left({3\beta_0\over 16}\left[{\pi^2\over 4}-1\right]\right)
\sim 20$
arises from the scale dependence of the strong coupling constant
in addition to the logarithmic dependence of the 
one-loop gluon graphs themselves.
Somewhat surprisingly, it is independent of $g_s$.
This factor will not be recovered by simply evaluating the leading
term at a renormalization point between $0$ and $2\mu$, as such corrections 
are of the form $\exp\left( r g_s \log\left(\lambda/\mu\right)\right)$,
where $r$ is some number.
The fourth factor is of order unity, and must be determined by matching
to a full QCD calculation, which we have not done.


To make sure the enhancement found above provides a good estimate of
the true effect of the running of $g_s$ we have numerically solved
eq.~(\ref{eq:gaplow}) for $\alpha_\Lambda\sim\overline{\alpha}$.
As the full QCD calculation has not been performed we set $C(2\mu,2\mu)
\sim C(2\mu,\alpha_\Lambda) =0$, since it is not enhanced by
$\beta_0$.  For $\mu=10^6~{\rm MeV}$ ($n_f=2$) and $\alpha_\Lambda =
\overline{\alpha}$ we find $\D=4.4~{\rm MeV}$ for $\beta_0=0$, and
$\D=32.5~{\rm MeV}$ for $\beta_0=11$.  For this density $\D$ is
enhanced by a factor of $\sim 7$ by the running of $g_s$.  Similarly,
for $\mu=10^4~{\rm MeV}$ and $\alpha_\Lambda = \overline{\alpha}$ we
find $\D=5.7~{\rm MeV}$ for $\beta_0=0$, and $\D=43.0~{\rm MeV}$ for
$\beta_0=11$. The difference between the estimated enhancement ($\sim 20$) and the
one found numerically ($\sim 7$) may be attributed  to the 
approximation in eq.(\ref{eq:brutality}). It is reassuring 
that $\D$ changes by only $\sim 2$
when the cut-off momentum is increased by an order of magnitude, while
keeping $C(2\mu,2\mu)
\sim C(2\mu,\alpha_\Lambda) =0$.  

In producing these
numbers the $q_0\ll |{\bf q}|$ limit of the gluon screening functions
was employed.  While we expect that such an approximation will modify
the results by a factor of order unity, we do not expect the
enhancement due to the running of $g_s$ to be substantially different.


In conclusion, we have found a significant enhancement in the size
of the color-superconducting gap, $\D$, due to the scale dependence
of the strong coupling constant, $g_s$.
We have resumed leading logarithms using the renormalization group
and obtained a scale independent gap equation to leading order in the 
effective field theory expansion in $g_s$ and $1/\mu$.
Terms that do not involve the large factor of $\beta_0$ have not
been calculated in this work, but are naively estimated to be smaller.

Matching to the effective theory needs to be performed at higher
orders to make a precise prediction for $\D$.  In our calculations, we
have assumed that both the magnetic and electric gluons have the same
$\beta$-function for all scales below $2\mu$.  Below the Debye mass,
the $\beta$-function for the magnetic gluons will deviate due to the
absence of logarithmic terms from electric gluon loops.  However, the
difference is expected to be of order $20\%$ and  our work
should be taken to be  only an estimate of this effect.

\vskip 1in

This work is supported in part by the U.S. Dept. of Energy under
Grants No. DE-FG03-97ER4014 and DOE-ER-40561.

\vfill\eject

\end{document}